\documentclass[reqno,12pt]{article}
\usepackage{amsmath,amsfonts,amssymb,amsthm,amstext,amscd,eucal,xcolor}
\usepackage[all]{xy}
\usepackage{mathtools}
\newcommand{\mathsym}[1]{{}} 
\usepackage[colorlinks=true,
            linkcolor=blue,
            urlcolor=blue,
            citecolor=blue]{hyperref}
\usepackage{enumerate}
\usepackage[utf8]{inputenc}

\usepackage{graphicx}
\usepackage{bm}
\usepackage{cite,hyperref}
\usepackage{mathtools}
\usepackage{graphicx}
\usepackage[active]{srcltx}
\makeatletter \@addtoreset{equation}{section}

\makeatletter\renewcommand\section{\@startsection {section}{1}{\z@}%
                                   {-3.5ex \@plus -1ex \@minus -.2ex}
                                   {2.3ex \@plus.2ex}%
                                   {\normalfont\large\bfseries}}
\renewcommand\subsection{\@startsection{subsection}{2}{\z@}%
                                     {-3.25ex\@plus -1ex \@minus -.2ex}%
                                     {1.5ex \@plus .2ex}%
                                     {\normalfont\bfseries}}

\DeclareMathAlphabet{\mathcal}{OMS}{cmsy}{b}{n}

\makeatother

\newcommand{\email}[1]{\footnote{E-mail: \href{mailto:#1}{#1}}}

\begin{document}

\title{\bf\Large{ Induced Maxwell-Chern-Simons Effective Action \\ in Very Special Relativity }}

\author{\textbf{R.~Bufalo\email{rodrigo.bufalo@ufla.br} $^{1}$, M.~Ghasemkhani\email{ghasemkhani@ipm.ir } $^{2}$, Z. Haghgouyan\email{z$_{_{-}}$haghgooyan@sbu.ac.ir} $^{2}$ and A.~Soto\email{arsoto1@uc.cl} $^{3}$ }\\\\
\textit{$^{1}$ \small Departamento de F\'isica, Universidade Federal de Lavras,}\\
\textit{ \small Caixa Postal 3037, 37200-000 Lavras, MG, Brazil}\\
\textit{\small $^{2}$ Department of Physics, Shahid Beheshti University, 1983969411, Tehran, Iran }\\
\textit{\small $^{3}$  Instituto de F\'isica, Pontificia Universidad de Cat\'olica de Chile,}\\
\textit{\small Av. Vicu\~na Mackenna 4860, Santiago, Chile}\\
}

\maketitle
\date{}
\begin{abstract}
In this paper, we study the one-loop induced photon's effective action in the very special relativity electrodynamics in $(2+1)$ spacetime (VSR-QED$_{3}$).
Due to the presence of new nonlocal couplings resulting from the VSR gauge symmetry, we have additional graphs contributing to the $\langle AA\rangle$ and $\langle AAA \rangle$ amplitudes.
From these contributions, we discuss the VSR generalization of the Abelian Maxwell-Chern-Simons Lagrangian, consisting in the dynamical part and the Chern-Simons-like self-couplings, respectively.
We use the VSR-Chern-Simons electrodynamics to discuss some non-Ohmic behavior on topological materials, in particular VSR effects on Hall's conductivity.
In the dynamical part of the effective action, we observe the presence of a UV/IR mixing, due to the entanglement of the VSR nonlocal effects to the quantum higher-derivative terms.
Furthermore, in the self-coupling aspect, we verify the validity of the Furry's
theorem in the VSR-QED$_{3}$ explicitly.

\end{abstract}


\setcounter{footnote}{0}
\renewcommand{\baselinestretch}{1.05}  

\newpage

\section{Introduction}

Over the past decades, we have seen great progress on the experimental and theoretical understanding of the standard model of particle physics and also cosmology, where high precision data have established tight bounds in search of manifestations of physics beyond the standard model \cite{ref53,Jacobson:2005bg,Bluhm:2005uj}.
Many minimal modifications of the standard model have been proposed and explored in order to understand the fundamental origin of some of the problems of physical phenomena that are not adequately explained by the known theories, e.g., neutrino masses, matter-antimatter asymmetry, quantum gravity, etc.
In general, a very reliable way to describe such phenomena involves the addition of new(global and/or local) degrees of freedom. However, instead of adding new fields, an appealing way to incorporate new degrees of freedom in this context is by enforcing a symmetry principle.

Among the enormous class of models that try to cope with physics beyond of the standard model, models presenting Lorentz symmetry violation have received considerable attention because they are usually related to physics at the Planck energy scale $E_{\rm Pl}$ \cite{ref53,AmelinoCamelia:2008qg}.
On the other hand, Lorentz violating effects are not necessarily related to Planck scale physics, it is also possible to formulate such class of models from a phenomenological group theory point of view.
In particular, the most interesting proposals are those that preserve the basic elements of special relativity, because they are in agreement with well-established physics.
A framework satisfying the above criteria is the Cohen and Glashow very special relativity (VSR) \cite{Cohen:2006ky,Cohen:2006ir}.
The main aspect of the VSR proposal is that the laws of physics
are invariant under the (kinematical) subgroups of the Poincar\'e group, preserving the basic elements of special relativity.
 Many interesting theoretical and phenomenological aspects of VSR effects have been extensively discussed  \cite{Dunn:2006xk,Alfaro:2015fha,Lee:2015tcc,Nayak:2016zed,Alfaro:2017umk,Alfaro:2019koq,Bufalo:2019qot}.

The kinematics of the VSR framework, when it is defined in $(3+1)$-dims, have two subgroups satisfying the prior requirements, namely, the HOM(2) (with three parameters) and the SIM(2) (with four parameters).
These symmetry groups SIM(2) and HOM(2) have the property of preserving the direction of a lightlike four-vector $n_{\mu}$ by scaling, transforming as $n \to e^{\varphi} n$ under boost in the z direction.
This feature implies that theories, which are invariant under either of these two subgroups, have a preferred direction in the Minkowski spacetime, where Lorentz violating terms can be constructed as ratios of contractions of the vector $n_{\mu}$ with other kinematical vectors \cite{Cohen:2006ky}.\footnote{The simplest example of VSR models is the free scalar field, whose action is given by
 $S = \int d^\omega x \tilde{\partial}_{\mu} \phi \tilde{\partial}^{\mu} \phi = \int d^\omega x \phi \left( -\Box +m^2 \right) \phi  $, where the wiggle derivative operator is defined by $\tilde{\partial}_{\mu}=\partial_{\mu}+\frac{1}{2}\frac{m^{2}}{n.\partial}n_{\mu}$.
We observe that the Lorentz violation appears in a nonlocal form and the parameter $m$ sets the scale for the VSR effects.}

Once the formulation of VSR is solely based on the use of kinematical subgroups of the Poincar\'e group, it can be
naturally extended to different spacetime dimensionality other than $(3+1)$-dims with proper considerations.
In the case of $(2+1)$-dims spacetime, we have in VSR the kinematical subgroup SIM$(1)$ subgroup (of the $SO(2,1)$ Lorentz group) that preserves all the aforementioned conditions,
in particular the existence of the invariant null-vector by a rescaling, and hence can be used to formulate a gauge theory in this framework \footnote{A detailed account of the SIM$(1)$ subgroup can be found in Ref.~\cite{ref51}.}.
In view of this thought, we are interested in formulating the Maxwell-Chern-Simons theory in the VSR context.
Since the proposal of the so-called topologically massive electrodynamics \cite{ref42},  also known as the Maxwell-Chern-Simons (MCS) electrodynamics, describing a single massive gauge mode of helicity $\pm1$, a great amount of attention has been paid into phenomenological application of this model.
From the theoretical point of view, an interesting aspect of 3D field theories is the UV finiteness in some models.
This feature, when is applied to Lorentz violating models, provides an ambiguity free description of Lorentz violation, allowing a close contact of violating effects with planar phenomena.
Furthermore, recently 3D versions of fermionization/bosonization have been studied \cite{refnew1,refnew2}, in which the duality between nonspin Chern-Simons theory and a spin Chern-Simons theory was investigated. Accordingly,
this work may be the first step in developing such analysis to VSR framework.

The dynamics of the gauge fields in the VSR framework has been studied in different applications \cite{Cheon:2009zx,Alfaro:2013uva,Bufalo:2016lfq,Alfaro:2019koq,Alfaro:2020njh}.
We shall discuss in this paper; however, the gauge fields dynamics in $(2+1)$-dims following the approach of effective action \cite{Niemi:1983rq}.
The approach of effective action allows a clear understanding of the low-energy dynamics of quantum fields,  where new types of interactions that depend on the spin of
the fields involved as well as the spacetime dimensionality can be obtained \cite{Bonora:2016otz,Bonora:2017ykb}.
Since VSR effects can be translated into the propagation of massive modes of the fields, we want to compare the possible difference of massive modes generation between the VSR effects and the Chern-Simons term due to the parity symmetry violation.
Furthermore, we highlight possible applications of the VSR-Chern-Simons effective action in the topological invariant systems, such as quantum Hall systems and other topological insulators.
We believe that  Lorentz violating effects can be well motivated in the context of topological invariant systems in order to describe anomalous behavior, such as non-Ohmic materials \cite{refpuica,refLo}.

In this paper, we examine in details the modifications of the photon's dynamics within the SIM$(1)$ VSR effective action framework.
Moreover, this analysis also allows us to verify the validity of Furry's theorem explicitly at one-loop order.
We start Sec.~\ref{sec2} by reviewing some aspects of the fermionic electrodynamics in the VSR context.
There, we explain the subtle points related to the nonlocal gauge couplings introduced by VSR effects, and also present the respective vertex Feynman rules.
We also discuss charge-conjugation symmetry in the Lagrangian level, showing that the C-invariance is preserved in the VSR setting.
In Sec.~\ref{sec3}, we compute the 2-point function $\langle AA \rangle$ at one-loop, corresponding to the dynamical part of the photon's effective action, and present the VSR modifications to the photon's polarization tensor.
In order to make contact with phenomenology, we discuss the VSR contribution, in terms of the VSR-Chern-Simons electrodynamics, to Hall's conductivity, as a possibility to describe non-Ohmic materials.
We analyze whether the VSR nonlocal couplings are sufficient to generate Chern-Simons-like self-couplings (in the Abelian VSR theory), and also check the Furry's theorem, by computing the 3-point function $\langle AAA \rangle$ in Sec. \ref{sec4}.
This discussion is well motivated since Lorentz invariance is one of the cornerstones of many fundamental and classical theorems and features in QFT. It is important to re-explore these theorems and features in the absence thereof. The VSR theories are among the few settings where these questions can be formulated and studied. Furry's theorem is one such example.
Finally, we summarize the results, and present our final remarks and prospects in Sec. \ref{conc}.


\section{Gauge fields in VSR}
\label{sec2}

In order to discuss the one-loop photon's effective action in VSR, we start by considering Dirac fermions interacting with an external gauge field as below
\begin{equation}
\mathcal{L}= \overline{\psi}\left[i\gamma^{\mu}\nabla_{\mu}-m_{e}\right]\psi.
\label{eq:1}
\end{equation}

The gauge and fermion fields are minimally coupled through the  VSR covariant derivative \cite{Alfaro:2013uva}
\begin{equation}
\nabla_{\mu}\psi=D_{\mu}\psi 	+\frac{1}{2} \frac{m^2}{\left(n. D\right)} n_{\mu}\psi, \label{eq:2a}
\end{equation}
which is written in terms of the ordinary covariant derivative $D_\mu = \partial_\mu -ie A_\mu $, and the preferred null direction is chosen as $n_{\mu}=\left(1,0,1\right)$. We observe that the expression \eqref{eq:2a} recovers the wiggle derivative $\tilde{\partial}_{\mu}=\partial_{\mu}+\frac{1}{2}\frac{m^{2}}{n.\partial}n_{\mu}$ in the noninteracting limit when $A_\mu \to 0$.
This new operator obeys the known transformation law for a charged field $\delta\left(\nabla_{\mu}\psi\right)=i\chi\left(\nabla_{\mu}\psi\right)$ and also satisfies
the required properties VSR gauge transformation $\delta A_{\mu}=\tilde{\partial}_{\mu}\chi$.

The fermionic propagator can readily be computed as
\begin{equation}
S(p)=\frac{i\left(\displaystyle{\not}\tilde{p}+m_{e}\right)}{\tilde{p}^{2}-m_{e}^{2}},
\label{eq:3}
\end{equation}
where $\tilde{p}_{\mu}=p_{\mu}-\frac{1}{2}\frac{m^{2}}{n.p}n_{\mu}$, and we also have the fermionic dispersion relation $\tilde{p}^{2}=m_{e}^{2}$, or equivalently as $p^{2}=\mu^{2}$, where $\mu^2 = m^{2}_{e} +m^2$ is the modified fermionic mass.

On the vertex functions, we should analyze the expression \eqref{eq:2a} for the VSR covariant derivative.
About the perturbative analysis, the presence of the term $1/\left(n.D\right)$ in \eqref{eq:2a} shows that there is now an infinite number of nonlocal interactions (in the coupling $e$).
The Feynman rules for these interactions can be obtained by means of the use of Wilson lines, which express the respective terms in a convenient form with $n=1,2,3,...$ legs of photon fields \cite{Dunn:2006xk}, making perturbative analysis workable.
Since we are interested in computing the vevs $\langle AA\rangle$ and $\langle AAA \rangle$ at one-loop order, we should consider the Feynman rules with one, two and three photon legs.
The $\left\langle \overline{\psi} \psi A\right\rangle $ vertex already signals minimal deviation from the usual QED, while the $\left\langle \overline{\psi} \psi AA\right\rangle $ and $\left\langle \overline{\psi} \psi AAA\right\rangle $ are new vertices resulting from VSR effects.
The 1PI vertex Feynman rules of interest can readily be obtained from the Lagrangian Eq.~\eqref{eq:1}\cite{Dunn:2006xk,Alfaro:2020njh}.
We would like to emphasize that the vertex Feynman rules are the same in any dimension, since the form of the interaction parts are preserved by changing the spacetime dimension.
However, the mass dimension of the fields and coupling constant depends directly on the spacetime dimension; e.g. the mass dimension of the fields and coupling constant in $d=2+1$ is described as $[\psi]=1$, $[A_{\mu}]=\frac{1}{2}$ and $[e]=\frac{1}{2}$, respectively.

\begin{figure}[t]
\vspace{-1.2cm}
\includegraphics[height=5\baselineskip]{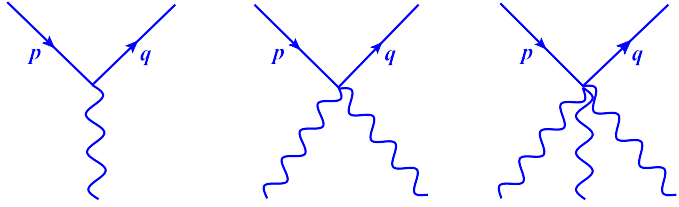}
 \centering\caption{ Feynman graphs of VSR-QED up to order $e^{3}$.}
\label{vertex_rules}
\end{figure}

By considering the three-dimensional Fermionic action as $S_{F}=\int d^{3}x~{\cal{L}}$, the relevant vertex Feynman rules, corresponding to the Feynman graphs depicted in Fig.~\ref{vertex_rules}, are found as below

\begin{itemize}
  \item The 3-point function $\langle \bar{\psi} (p) \psi(q) A_\mu (k)\rangle $
\begin{align}
\Lambda^{\mu}   & =-ie\left[\gamma^{\mu}+\frac{m^{2}}{2}\frac{\displaystyle{\not}n~n^{\mu}}
{\left(n.p\right)\left(n.q\right)}\right]
\label{eq:4}
\end{align}
  \item The 4-point function $\langle \bar{\psi} (p) \psi(q) A_\mu (k_1)A_\nu (k_2)\rangle $
\begin{equation}
\Gamma^{\mu\nu}  =-\frac{ie^{2}m^{2}}{2}\frac{\displaystyle{\not}n~n^{\mu}n^{\nu}}{(n.p)(n.q)}\left[\frac{1}{n.(p+k_1)}+\frac{1}{n.(p+k_2)}\right]
 \label{eq:5}
\end{equation}
  \item The 5-point function $\langle \bar{\psi} (p) \psi(q) A_\mu (k_1)A_\nu (k_2)A_\rho (k_3)\rangle $
\begin{equation}
\Gamma^{\mu\nu\rho} =-\frac{ie^{3}m^{2}}{2}\frac{\displaystyle{\not}n~n^{\mu}n^{\nu}n^{\rho}}{(n.p)(n.q)}\left[ \Big(\frac{1}{n.\left(p+k_1\right) }+\frac{1}{n.\left(p+k_2\right) } \Big)\frac{1}{ n.\left(p+k_1+k_2\right)} + {\rm perm}\right]
 \label{eq:6}
\end{equation}
\end{itemize}

In these vertex functions, we have assumed that all the photon momenta are inward, implying the energy-momentum conservation law $p-q+\sum\limits_i k_i=0$.

In regard to the discrete symmetries in this Lorentz violating setting, one should recall that the VSR context does not permit the inclusion of parity (P) and time-reversal (T) symmetries, as well as composed symmetries CP and CT, since these are sufficient to restore the full Lorentz invariance \cite{Cohen:2006ky}.
On the other hand, we can speak about the charge-conjugation symmetry in regard to the nonlocal VSR couplings, because it does not depend on the structure of the spacetime.
One of the main goals in this paper is investigating the validity of Furry's theorem in the VSR setting. And before proceeding to the analysis of the vev $\langle AAA \rangle$ at one-loop order, we shall first study the charge-conjugation symmetry at the Lagrangian density (classical level).
With this purpose, we first consider the behavior of the free part of the Lagrangian \eqref{eq:1} under C transformation which is given by
\begin{equation}
{\cal{L}}_{0}^{c}=\bar\psi^{c}\Big[i\displaystyle{\not}\partial-m_{e}
+i\frac{m^{2}}{2}\frac{\displaystyle{\not}n}{n.\partial}\Big]\psi^{c},
\end{equation}
where $\psi^{c}=\eta_{_{\psi}}C\bar\psi^{T}$ and $\bar\psi^{c}=-\eta^{*}_{_{\psi}}\psi^{T}C^{-1}$.
Using the anticommuting property of the spinor components and the identity $C^{-1}\gamma^{\mu}C=-(\gamma^{\mu})^{T}$, it is easy to identify that the first and second terms are C-invariant, by a total derivative term in the action.
About the last term, including the VSR effects, it transforms as
\begin{equation}
\bar\psi^{c}\gamma^\mu \left(\frac{1}{n.\partial}\psi^{c} \right)
=-\Big(\frac{1}{n.\partial}\bar\psi\Big) \gamma^\mu\psi.
\end{equation}
With the help of the identity $\frac{1}{n.\partial}=\int_{0}^{\infty}ds~e^{-s(n.\partial)}$, it is easy to show that under the action integral
\begin{equation}
\int d^{3}x~\bar\psi^{c}\gamma^\mu \left(\frac{1}{n.\partial}\psi^{c} \right)
=-\int d^{3}x\Big(\frac{1}{n.\partial}\bar\psi\Big)\gamma^\mu\psi=
\int d^{3}x~\bar\psi \gamma^\mu \left(\frac{1}{n.\partial}\psi \right).
\end{equation}
Finally, with this result, we can conclude that the free part of the action ${\cal{S}}_{0}=\int d^{3}x~{\cal{L}}_{0}$ is C-invariant.
The remaining part of the analysis consists of checking the behavior of the interaction part of the Lagrangian \eqref{eq:1} under C.
It is well known that the usual QED interaction term, i.e. $\bar\psi\gamma^{\mu}A_{\mu}\psi$, is explicitly C-invariant.
However, the  nonlocal interaction terms, arising from VSR effects, are generated by the perturbative expansion of the term
\begin{equation}
\bar\psi\gamma^\mu \left(\frac{1}{n.D}\psi \right)=\bar\psi \gamma^\mu \left( \int_{0}^{\infty}ds~e^{-s(n.D)}\psi \right).
\end{equation}
Similarly to the analysis of the free part under C, and by considering $A_{\mu}^{c}=-A_{\mu}$, it is straightforward to show that the whole VSR (nonlocal) couplings will be C-invariant under the action integral, added by total derivatives.
Hence, we can conclude that C-invariance is preserved in the VSR setting at the classical level.
We shall return to the C-invariance in Sec.~\ref{sec4} through the explicit verification of the Furry's theorem at one-loop order in the VSR context.

\section{One-loop 2-point function $\langle A_\mu A_\nu \rangle$}
\label{sec3}

\begin{figure}[t]
\vspace{-1.2cm}
\includegraphics[height=6\baselineskip]{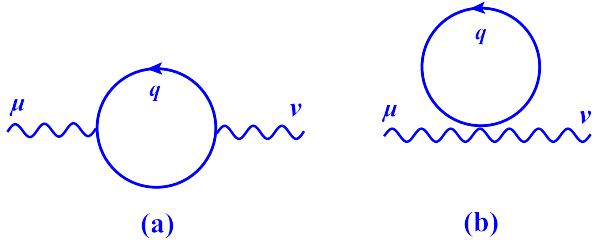}
 \centering\caption{Relevant graphs for the induced $AA$-term.}
\label{oneloop1}
\end{figure}

In order to compute the dynamical part of the photon's effective action, we shall discuss the 2-point function $\langle A_\mu A_\nu \rangle$ at one-loop order.
The Feynman diagrams of the respective contributions are depicted in Fig.~\ref{oneloop1}.
The first graph (a) corresponds to the usual photon polarization of QED, where the fermion propagator \eqref{eq:3} and vertex \eqref{eq:4} are modified by VSR nonlocal terms, resulting in
\begin{equation}
\Pi_{(a)}^{\mu\nu}\left(p\right)=-\int\frac{d^{\omega}q}{\left(2\pi\right)^{\omega}}\textrm{Tr}
\left(i\frac{\left(\displaystyle{\not}\tilde{q}+m_{e}\right)}{\tilde{q}^{2}-m_{e}^{2}}\Lambda^{\mu}\left(q,p+q\right)i\frac{\left(\displaystyle{\not}\tilde{u} +m_{e}\right)}{ \tilde{u}^{2}-m_{e}^{2}}\Lambda^{\nu}\left(p+q,q\right)\right),
\end{equation}
in which $u = p+q$. There is a second VSR contribution at this order, graph (b), corresponding to the quartic VSR vertex \eqref{eq:5},  which gives
\begin{equation}
\Pi_{(b)}^{\mu\nu}\left(p\right)=-\int\frac{d^{\omega}q}{\left(2\pi\right)^{\omega}}
\textrm{Tr}\left(i\frac{\left(\displaystyle{\not}\tilde{q}+m_{e}\right)}{\tilde{q}^{2}-m_{e}^{2}}
\Gamma_{\mu\nu}\left(q,-q,p,-p\right)\right).
\end{equation}
Then, the full contribution to the $\langle A_\mu A_\nu \rangle$ part is given by $\Pi^{\mu \nu} = \Pi_{(a)}^{\mu\nu}+\Pi_{(b)}^{\mu\nu}$.
In the first part of the computation, we make use of the known algebra for the Dirac's gamma matrices in the two-component representation in the $(2+1)$ spacetime,
\[
\textrm{Tr}\left(\gamma^{\sigma}\gamma^{\rho}\right)=2\eta^{\sigma\rho},
\quad\textrm{Tr}\left(\gamma^{\alpha}\gamma^{\sigma}\gamma^{\rho}\right)=-2i\epsilon^{\alpha\sigma\rho},
\quad\textrm{Tr}\left(\gamma^{\alpha}\gamma^{\sigma}\gamma^{\lambda}\gamma^{\rho}\right)=2\left(\eta^{\alpha\sigma}
\eta^{\lambda\rho}-\eta^{\alpha\lambda}\eta^{\sigma\rho}+\eta^{\alpha\rho}\eta^{\lambda\sigma}\right).
\]
After simplifying the trace part of the polarization tensor, we obtain
\begin{align}
\Pi^{\mu \nu} \left(p\right) &= - 2 e^2 \int \frac{d^{\omega}q}{(2 \pi)^\omega}~\frac{ \tilde{q }^{\sigma}\tilde{u }^{\rho}-   \eta^{\sigma \rho} \left(\tilde{q }.\tilde{u }\right) +  \tilde{q }^{\rho}\tilde{u }^{\sigma}  }{(q^2 - \mu^2)
((p + q)^2 - \mu^2)}~  \mathcal{T}_{\sigma}^{\mu} (p,q)~  \mathcal{T}_{\rho}^{\nu} (p,q)   \cr
&+ 2 i e^2 m_e \int \frac{d^{\omega}q}{(2 \pi)^\omega}~\frac{ \varepsilon^{\alpha \sigma \rho}\tilde{q }_{\alpha} +  \varepsilon^{\sigma \beta \rho} \tilde{u }_{\beta}}{(q^2 - \mu^2) ((p + q)^2
- \mu^2)} ~ \mathcal{T}_{\sigma}^{\mu} (p,q) ~  \mathcal{T}_{\rho}^{\nu} (p,q)  \cr
&- 2 e^2 m_e^2 \int \frac{d^{\omega}q}{(2 \pi)^\omega}~\frac{1}{(q^2 - \mu^2) ((p + q)^2
- \mu^2)}   ~  \mathcal{T}^{\rho \mu} (p,q)   ~  \mathcal{T}_{\rho}^{\nu} (p,q)  \cr
&- e^2 m^2 n^{\mu} n^{\nu} \int \frac{d^{\omega}q}{(2 \pi)^\omega}~\frac{1}{q^2 - \mu^2}
\frac{1}{(n.q)} \Big[\frac{1}{n.(q + p)} + \frac{1}{n.(q - p)} \Big],
\label{eq:7}
\end{align}
where we have defined by notation the nonlocal tensor as
\begin{equation}
\mathcal{T} _{\alpha\beta} (p,q)=\eta_{\alpha\beta} + \frac{m^2}{2} \frac{n_{\alpha} n_{\beta}}{ (n. q) n.(p + q)}.
\end{equation}
In order to solve the momentum integrals, it is convenient to use the following property
 \begin{equation}
 \frac{1}{ n.\left(q+p_i\right)n.\left(q+p_j\right)} = \frac{1}{ n.\left(p_i-p_j\right)} \left(  \frac{1}{ n.\left(q+p_j\right)} - \frac{1}{ n.\left(q+p_i\right)}\right),
  \label{eq:11}
 \end{equation}
and simplify the integrand of equation  \eqref{eq:7}.
In the case of infrared divergent integrals, we can use the Mandelstam-Leibbrandt prescription\cite{Mandelstam:1982cb, Leibbrandt:1983pj}
\begin{equation}
\frac{1}{(n.q)} = \lim_{\varepsilon \to 0} \frac{(q.\bar{n})}{(n.q)(q.\bar{n})+i\varepsilon},
\end{equation}
where $\bar{n}$ is a new null vector with the property $(n.\bar{n}) = 1 $, whose explicit form must be provided in the VSR framework \cite{Alfaro:2017umk}.
We discuss the conditions to determine the specific form of $\bar{n}$ below.
One main result for the momentum integrals involved with the $(n\cdot q)^{-1}$ factor is \cite{Alfaro:2016pjw,Alfaro:2017umk}
 \begin{align}
\int d^{\omega}q \frac{1}{(q^2 + 2 q.p - m^2)^a} \frac{1}{(n. q)^b}
 = (- 1)^{a + b} i \pi^{\frac{\omega}{2}} (- 2)^b \frac{\Gamma (a + b - \frac{\omega}{2})}{\Gamma
(a) \Gamma (b)} (\bar{n}.p)^b \int^1_0 d t~t^{b - 1} \frac{1}{ \Delta^{a + b - \frac{\omega}{2}}},
\label{eq:8}
\end{align}
where $\Delta = m^2 + p^2 - 2 (n.p) (\bar{n}.p) t$.
Hence, making use of the identity \eqref{eq:11} and inserting back the results Eqs.~\eqref{eq:8}, \eqref{eq:9} and \eqref{eq:10}, into the expression \eqref{eq:7}, it yields
\begin{align}
\Pi^{\mu\nu}(p)&=\frac{4ie^{2}}{(4\pi)^{\frac{\omega}{2}}}\Gamma\left(2-\frac{\omega}{2}\right)
\left(p^{\mu}p^{\nu}-\eta^{\mu\nu}p^{2}\right)\int_{0}^{1}dx~x(1-x)\Big(\frac{1}{\mu^{2}-x(1-x)p^{2}}\Big)^{2-\frac{\omega}{2}}\cr
&+\frac{2e^{2}m_{e}}{(4\pi)^{\frac{\omega}{2}}}\Gamma\Big(2-\frac{\omega}{2}\Big)\varepsilon^{\alpha\mu\nu}p_{\alpha}\int_{0}^{1}dx\Big(\frac{1}{\mu^{2}-x(1-x)p^{2}}\Big)^{2-\frac{\omega}{2}} \label{eq:12}\\
&+\frac{4ie^{2}m^{2}}{(4\pi)^{\frac{\omega}{2}}}\Gamma\left(3-\frac{\omega}{2}\right)\left[n^{\mu}p^{\nu}+n^{\nu}p^{\mu}-\eta^{\mu\nu}\left(n. p\right)-\frac{n^{\mu}n^{\nu}}{\left(n.p\right)}p^{2}\right]\left(\bar{n}. p\right)\int_{0}^{1}dx~x\int_{0}^{1}dt\frac{1}{\varTheta^{3-\frac{\omega}{2}}}\cr
&+\frac{4e^{2}m^{2}m_{e}}{(4\pi)^{\frac{\omega}{2}}}\Gamma\left(3-\frac{\omega}{2}\right)\left[\varepsilon^{\alpha\sigma\nu}p_{\alpha}\frac{n_{\sigma}n^{\mu}}{n. p}+\varepsilon^{\alpha\mu\rho}p_{\alpha}\frac{n_{\rho}n^{\nu}}{n.p}+\varepsilon^{\alpha\mu\nu}n_{\alpha}\right]\left(\bar{n}. p\right)\int_{0}^{1}dx~x\int_{0}^{1}dt\frac{1}{\varTheta^{3-\frac{\omega}{2}}}, \nonumber
\end{align}
where $\varTheta=\mu^{2}-x(1-x)p^{2}-2x^{2}(n.p)(\bar{n}. p)t$.
We can explicitly observe that the Mandelstam-Leibbrandt prescription preserves the VSR gauge invariance, since \eqref{eq:12} satisfies the Ward identity $p_{\mu}\Pi^{\mu \nu}  =p_{\nu}\Pi^{\mu \nu}  = 0$.

Now, in order to fully determine the integrals in \eqref{eq:12}, we must present an explicit form for the vector $\bar{n}$.
Taking into account the properties such as
reality, right scaling $(n,\bar{n}) \to (\lambda n, \lambda^{-1}\bar{n})$  and being dimensionless \cite{Alfaro:2017umk}, we find a SIM(1)-invariant vector $\bar{n}_\mu = a  \frac{p^2 n_\mu}{ (n.p)^2}  +b \frac{p_\mu}{(n.p)} +c \frac{\sqrt{q^2} \epsilon_{\mu \rho \sigma} n^\rho p^\mu}{(n.p)^2}$, with $a,b,c$ pure numbers.
Notice, however, that this fails to be real for $q^2<0$.
Thus, to preserve reality, it is necessary to consider $c=0$.
Finally, in our prescription, we have that $\bar{n}_{\mu} = - \frac{p^2}{2 (n.p)^2} n_{\mu}
+ \frac{p_{\mu}}{n.p}$, then $\bar{n}.p = \frac{p^2}{2 (n.p)}$.
Hence, replacing this result back into \eqref{eq:12}, solving the integration over the variable $t$, we arrive at
\begin{align}
\Pi^{\mu \nu}(p)&= \frac{ie^2}{2 \pi}  \left(p^{\mu} p^{\nu} - \eta^{\mu \nu} p^2\right)
\mathcal{I}_1  + \frac{e^2}{4 \pi}  m_e \varepsilon^{\alpha \mu \nu} p_{\alpha} \mathcal{I}_2 \cr
&+ \frac{i e^2 m^2}{4 \pi}  \left[ \frac{n^{\mu} p^{\nu} + n^{\nu} p^{\mu}}{n.p} - \eta^{\mu \nu} - \frac{n^{\mu} n^{\nu}}{(n.p)^2} p^2
\right] \mathcal{I}_3 \cr
&+ \frac{e^2 m_e m^2}{4 \pi}  \left[ \varepsilon^{\alpha \sigma \nu}
p_{\alpha} \frac{n_{\sigma} n^{\mu}}{(n.p)^2} + \varepsilon^{\alpha
\mu \sigma} p_{\alpha} \frac{n_{\sigma} n^{\nu}}{(n.p)^2} +
\varepsilon^{\alpha \mu \nu} \frac{n_{\alpha}}{n.p} \right] \mathcal{I}_3,
 \label{eq:14}
\end{align}
where we have considered $\omega \to 3^+$, since the expression \eqref{eq:12} is UV finite in this limit, and we have also defined the integrals $\mathcal{I}_i$ by simplicity in Eqs.~\eqref{int_1}, \eqref{int_2} and \eqref{int_3}.

In order to determine the VSR contributions to the Maxwell-Chern-Simons kinetic terms for the photon's effective action, we consider the low-momentum limit $p^{2} \ll m_e^{2}$ of the expression \eqref{eq:14}.
In this case, the integrals behave as
\begin{align}
 \mathcal{I}_{1} \simeq  &\frac{1}{6 |\mu|}, \quad
\mathcal{I}_{2}  \simeq \frac{1}{|\mu|}, \quad
\mathcal{I}_{3} \simeq \frac{p^{2}}{4|\mu|^{3}},
  \label{eq:16}
\end{align}
where the higher-derivative corrections to the ordinary Maxwell-Chern-Simons term have been discarded.
Finally, the polarization tensor, in the low-momentum limit $p^{2} \ll m_e^{2}$, can be written in the form
\begin{align}
-i\Pi^{\mu\nu}(p)\bigg|_{p^{2} \ll m_{e}^{2}}&=\frac{e^{2}}{12\pi |m_{e}|}(p^{\mu}p^{\nu}-\eta^{\mu\nu} p^{2})
-\frac{ie^{2}m_e}{4\pi |m_e|}\epsilon^{\mu\nu\alpha}p_{\alpha}\cr
&+\frac{e^{2}}{16\pi | m_{e}|} \left(\frac{m^{2}}{m^{2}_{e}}\right)\Big[\frac{n^{\mu}p^{\nu}+n^{\nu}p^{\mu}}{n.p}-\eta^{\mu\nu}-\frac{n^{\mu}n^{\nu}}{(n.p)^{2}}p^{2}\Big]
p^{2}\cr
&-\frac{ie^{2} m_e}{16\pi |m_e|}  \left(\frac{m^{2}}{m^{2}_{e}}\right)\Big[\epsilon^{\mu\nu\alpha}\frac{n_{\alpha}}{n.p}+\epsilon^{\alpha\sigma\nu}p_{\alpha}
\frac{n^{\mu}n_{\sigma}}{(n.p)^{2}}+\epsilon^{\alpha\mu\sigma}p_{\alpha}\frac{n^{\nu}n_{\sigma}}{(n.p)^{2}}\Big]
p^{2}.
 \label{eq:15}
\end{align}
The low-energy VSR photon's effective action is obtained by using the result \eqref{eq:15} into
\begin{equation}
i\Gamma [A] = \int \frac{d^3p}{(2\pi)^3}\int d^3x_1 d^3x_2 e^{ip(x_1-x_2)} A_{\mu}(x_1) A_{\nu}(x_2) \Pi^{\mu\nu}(p)\bigg|_{p^{2} \ll m_{e}^{2}}.
\end{equation}
yielding the following induced Lagrangian density
\begin{align} \label{eq:15.1}
\mathcal{L}_{\tiny\mbox{ind.}} & =-\frac{1}{4 }F^{\mu\nu}F_{\mu\nu}
 +  \frac{3 m_e}{4  }\epsilon^{\mu\nu\alpha}A_{\mu}F_{\nu\alpha}
\nonumber \\
 & - \frac{3 m^2}{8 }
\left(n_{\nu}F^{\mu\nu}\right)\frac{1}{\left(n.\partial\right)^{2}}  \left(\frac{\Box}{m^{2}_{e}}\right) \left(n^{\alpha} F_{\mu\alpha}\right)
 \nonumber\\
  & - \frac{ 3 m^2 m_e }{16 }
    \Big[ 2\epsilon^{\mu\nu\lambda}A_{\lambda}
 \frac{n_{\nu}n^{\alpha}}{\left(n.\partial\right)^{2}}  \left(\frac{\Box}{m^{2}_{e}}\right) F_{\mu\alpha}+
 \epsilon^{\mu\alpha\nu}A_{\lambda}\frac{n_{\nu}n^{\lambda}}{\left(n.\partial\right)^{2}} \left(\frac{\Box}{m^{2}_{e}}\right) F_{\mu\alpha}\Big].
\end{align}
where, $\Gamma[A]= \left( \frac{e^2}{6\pi |m_e|}\right)\int d^{3}x~\mathcal{L}_{\tiny\mbox{ind.}}$ and $F_{\mu\nu}=\partial_{\mu}A_{\nu}-\partial_{\nu}A_{\mu}$.
As an illustration, we can compare the tensor structure of the  induced Lagrangian density Eq.~\eqref{eq:15.1}, to the classical Lagrangian, which is a VSR gauge invariant generalization of the ordinary Maxwell-Chern-Simons Lagrangian density
\begin{align} \label{eq:115}
\mathcal{L}_{2+1} & =-\frac{1}{4}F^{\mu\nu}F_{\mu\nu}-\frac{m^{2}}{2}\left(n_{\nu}F^{\mu\nu}\right)\frac{1}{\left(n.\partial\right)^{2}}\left(n^{\alpha}F_{\mu\alpha}\right)\nonumber \\
 & +\frac{m_{e}}{4}\epsilon^{\mu\nu\lambda}F_{\mu\nu}A_{\lambda}+\frac{m^{2}m_{e}}{4}\epsilon^{\mu\nu\lambda}
 \frac{n_{\nu}n^{\alpha}}{\left(n.\partial\right)^{2}}F_{\mu\alpha}A_{\lambda}+\frac{m^{2}m_{e}}{8}
 \epsilon^{\mu\alpha\nu}\frac{n_{\nu}n^{\lambda}}{\left(n.\partial\right)^{2}}F_{\mu\alpha}A_{\lambda}.
\end{align}
Furthermore, we can cast the Lagrangian \eqref{eq:115} as $\mathcal{L}_{2+1}  =\frac{1}{2}A_{\lambda}\mathcal{O}^{\lambda\alpha}A_{\alpha}$, where $\mathcal{O}^{\lambda\alpha}$ in the momentum space is given by
\begin{align} \label{eq:116}
\widetilde{\mathcal{O}}^{\lambda\alpha} & =k^{\lambda}k^{\alpha}-\eta^{\lambda\alpha}k^{2}+im_{e}\epsilon^{\mu\alpha\lambda}k_{\mu}\nonumber \\
 & +m^{2}\left(\frac{n^{\lambda}n^{\alpha}k^{2}}{\left(n.k\right)^{2}}-\frac{n^{\lambda}k^{\alpha}+n^{\alpha}k^{\lambda}}{\left(n.k\right)}+\eta^{\alpha\lambda}\right)\nonumber \\
 & +im_{e}\frac{m^{2}}{2}\left[\frac{\epsilon^{\alpha\nu\lambda}n_{\nu}}{\left(n.k\right)}
 -\frac{\epsilon^{\mu\nu\lambda}k_{\mu}n_{\nu}n^{\alpha}}{\left(n.k\right)^{2}}
 +\frac{\epsilon^{\mu\nu\alpha}k_{\mu}n_{\nu}n^{\lambda}}{\left(n.k\right)^{2}}\right].
\end{align}
We can easily observe that the quantum effective action \eqref{eq:15} and the classical action  \eqref{eq:116} have the same tensor structure in the VSR framework, showing that both VSR and gauge symmetry are preserved.
However, there is a major difference between the results, the VSR effects in the quantum counterpart  \eqref{eq:15} all come as higher-derivative terms.

It is well known that the infrared fluctuations can generate nonlocal terms in the quantum effective action.
This generation of the nonlocal terms can be traced back to the presence of massless particles in the fundamental theory, this is the case for example of QED and gravity \cite{Belgacem:2017cqo}.
On the other hand, higher derivative terms can also  be related to quantum effects \cite{Borges:2019gpz}.
Hence, from this point of view, we understand that in the VSR framework, the nonlocal and higher-derivative terms are intermingled  in the quantum effective action \eqref{eq:15}.

It is important to emphasize that this statement is not valid in the $(2+1)$ VSR electrodynamics only. In fact, if we analyze the generation of VSR gauge terms in the $(3+1)$ spacetime \cite{Alfaro:2017umk}  and $(1+1)$ spacetime \cite{Alfaro:2019snr}, we observe that all of the nonlocal effects are entangled to higher-derivative terms.
Hence, this mixture of nonlocal effects and higher-derivative terms in VSR, in the gauge sector of quantum effective action, seems to be a common feature.
Thus, we can signal that we have the presence of UV/IR mixing in the VSR quantum effective action.

\subsection{Topological insulators in VSR}

Chern-Simons theories are known to capture the response of the quantum Hall ground state to low-energy perturbations, more precisely it captures the basic physical content of the integer quantum Hall effect \cite{Tong:2016kpv}.
Moreover, systems exhibiting the quantum Hall effect were the first topological insulators as being characterized by a nonzero Chern number.
Within this class of systems, there are some cases that present a non-Ohmic behavior at high-temperature \cite{refpuica,refLo}.
From this point of view, one can use generalizations of the Chern-Simons theory in order to describe this anomalous behavior.
Hence, we wish to explore these rich environments in order to highlight the VSR effects, for this matter we consider the VSR-Chern-Simons electrodynamics \eqref{eq:115}.

Let us consider the Chern-Simons part of the  Eq.~\eqref{eq:115}, and applying the scaling $eA_{\mu}\rightarrow {\cal {A}}_{\mu}$, we have
 \begin{align}\label{eq:116-1}
{\cal{L}}_{_{\textrm{VSR-CS}}} =
 \frac{\kappa}{4}\epsilon^{\mu\nu\lambda}{\cal{F}}_{\mu\nu}{\cal{A}}_{\lambda}+\frac{m^{2}\kappa}{4}\epsilon^{\mu\nu\lambda}
 \frac{n_{\nu}n^{\alpha}}{\left(n.\partial\right)^{2}}{\cal{F}}_{\mu\alpha}{\cal{A}}_{\lambda}+\frac{m^{2}\kappa}{8}
 \epsilon^{\mu\alpha\nu}\frac{n_{\nu}n^{\lambda}}{\left(n.\partial\right)^{2}}{\cal{F}}_{\mu\alpha}{\cal{A}}_{\lambda},
\end{align}
and we define the effective action $S_{_{\textrm{VSR-CS}}}=\int d^{3}x~ {\cal{L}}_{_{\textrm{VSR-CS}}}$, with the parameter $\kappa=\frac{m_{e}}{e^{2}}$, which is a dimensionless quantity.

Hence, in order to study the non-Ohmic effects in the Hall's conductivity, we compute the mean current density, subject to the presence of an external field, as below
\begin{equation}\label{eq:116-2}
\left\langle J^{\sigma}\left(x\right)\right\rangle =\frac{\delta S_{_{\textrm{VSR-CS}}}\left[{\cal{A}}\right]}{\delta {\cal{A}}_{\sigma}\left(x\right)}.
\end{equation}
Then, substituting \eqref{eq:116-1} into \eqref{eq:116-2} yields us the expression
\begin{align}
\left\langle J^{\sigma}\left(x\right)\right\rangle  =\frac{\kappa}{2}\epsilon^{\sigma\mu\nu}{\cal{F}}_{\mu\nu}
 +\frac{m^{2}\kappa}{4}\left(\epsilon^{\mu\lambda\nu}n^{\sigma}+2\epsilon^{\sigma\mu\nu}n^{\lambda}\right)
 \frac{n_{\nu}}{\left(n.\partial\right)^{2}}{\cal{F}}_{\mu\lambda}.
\end{align}
Let us consider separately the cases $\sigma=0$ and $\sigma=i$, corresponding to the charge density and vector current density respectively. Thus, we find that
\begin{align}
\left\langle J^{0}\left(x\right)\right\rangle =\frac{\kappa}{2}\epsilon^{ij}{\cal{F}}_{ij} +\frac{m^{2}\kappa}{4}\frac{1}{(n.\partial)^{2}}{\cal{G}}^{(0)},
\end{align}
and
\begin{align}
\left\langle J^{i}\left(x\right)\right\rangle  =\frac{\kappa}{2}\epsilon^{i\mu\nu}{\cal{F}}_{\mu\nu}+\frac{m^{2}\kappa}{4}\frac{1}{\left(n.\partial\right)^{2}}{\cal{G}}^{(i)},
\end{align}
where we have defined by simplicity
\begin{align}
{\cal{G}}^{(0)}&=\epsilon^{\mu\lambda\nu}n^{0}n_{\nu}{\cal{F}}_{\mu\lambda}+2\epsilon^{0\mu\nu}n^{\lambda}n_{\nu}{\cal{F}}_{\mu\lambda},\cr
{\cal{G}}^{(i)}&=\epsilon^{\mu\lambda\nu}n^{i}n_{\nu}{\cal{F}}_{\mu\lambda}+2\epsilon^{i\mu\nu}n^{\lambda}n_{\nu}{\cal{F}}_{\mu\lambda}.
\end{align}

Finally, using explicitly the definition of $n_{\mu}=(1,0,1)$, it is straightforward to show that the VSR contributions to the mean current vanish, ${\cal{G}}^{(0)}={\cal{G}}^{(i)}=0$.
It shows that the Hall's conductivity $\sigma$ does not receive any VSR corrections.
Therefore, we are left with the usual relation among the charge and current densities in terms of the electric and magnetic fields
 \begin{align}\label{eq:hall}
\left\langle J^{0}\left(x\right)\right\rangle=\kappa{\cal{B}},\quad\quad\left\langle J^{i}\left(x\right)\right\rangle=\kappa\epsilon^{ij}{\cal{E}}_{j},
\end{align}
 with ${\cal{B}}=\frac{1}{2}\epsilon^{ij} {\cal{F}}_{ij}, {\cal{E}}^{i}={\cal{F}}^{i0}$.
By comparison, \eqref{eq:hall} corresponds to the expected Ohmic behavior described by the Chern-Simons theory for the Hall's conductivity with the identification $\kappa = e^2 n/\hbar $, where $n$ is the occupancy number of the Landau levels.
Although it was shown that the VSR effects vanishes, in this particular class of Hall's systems, there are still a considerable amount of topological invariant physical systems where VSR could be formulated in order to  describe anomalous behavior.
As discussed above, there are several different physical systems (topological insulators) characterized not only by the Chern number, but also by the Chern parity.
Hence, we believe that Lorentz violating effects could also be applied to these physical systems in order to discuss non-Ohmic behavior, or any other anomalous behavior, experimentally observed in some materials.


\section{One-loop 3-point function $\langle A_\mu A_\nu A_\sigma \rangle$}
\label{sec4}

\begin{figure}[t]
\vspace{-1.2cm}
\includegraphics[height=8\baselineskip]{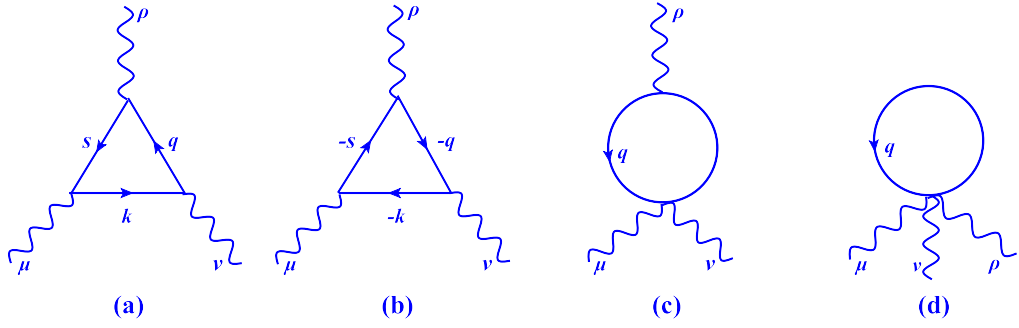}
 \centering\caption{Relevant graphs for the induced $AAA$-term.}
\label{oneloop2}
\end{figure}

A strong result in the standard QED is the C invariance known as the Furry's theorem.
This theorem states that the total amplitude of the graphs containing a closed fermion loop with an odd number of external photon legs vanishes.
Although the VSR electrodynamics \eqref{eq:1} is also invariant under charge conjugation, there are a number of additional couplings in comparison to QED, making necessary a   detailed analysis of the contribution of these couplings.
Hence, in this section, we shall verify explicitly whether Furry's theorem is valid at one-loop order in the presence of the nonlocal VSR couplings.
This analysis is also motivated by the fact that not necessarily such cornerstones of QFT are valid in a Lorentz violating setting, because such violating effects may result into an anomaly.
The four contributing graphs are depicted in Fig.~\ref{oneloop2}.
We observe that the triangle graphs (a) and (b) have the same structure as ordinary QED, whereas graphs (c) and (d) include purely VSR effects, coming from the new vertices \eqref{eq:5} and \eqref{eq:6}, respectively.

In order to verify whether the total amplitude, consisting of the diagrams in Fig.~\ref{oneloop2}, vanishes, we start by discussing the triangle graphs (a) and (b).
Following the set of Feynman rules for the model, we have
\begin{align}
\Xi^{\mu\nu\rho}_{ (a)}&=(-1) \int\frac{d^{\omega}q}{(2\pi)^{\omega}}
{\rm Tr}\Bigg[\frac{i(\displaystyle{\not}\tilde{k}+m_{e} )}{k^{2}-\mu^{2}}
\Lambda^{\nu}\left(k,q\right)  \frac{i\left(\displaystyle{\not}\tilde{q}+m_{e} \right)}{q^{2}-\mu^{2}}
\Lambda^{\rho}\left(q,s\right)
\frac{i\left(\displaystyle{\not}\tilde{s}+m_{e} \right)}{s^{2}-\mu^{2}}
\Lambda^{\mu}\left(s,k\right)\Bigg]
\end{align}
where, the external legs are denoted by the inward momenta $(p_{1}^\mu, p_{2}^\nu, p_{3}^\rho)$, and $k=q-p_{2}$, $s=q+p_{3}$. Moreover, the second graph (b) is given by
\begin{align}
\Xi^{\mu\rho\nu}_{ (b)} =&(-1)\int\frac{d^{\omega}q}{(2\pi)^{\omega}}\textrm{Tr}\Bigg[\frac{i\left(-{  \displaystyle{\not}\tilde{s}+m_{e}}\right)}{s^{2}-\mu^{2}}\Lambda^{\rho}\left(-s,-q\right)\frac{i\left(-{  \displaystyle{\not}\tilde{q}+m_{e}}\right)}{q^{2}-\mu^{2}} \Lambda^{\nu}\left(-q,-k\right) \frac{i(-{  \displaystyle{\not}\tilde{k}+m_{e}})}{k^{2}-\mu^{2}}\Lambda^{\mu}\left(-k,-s\right)\Bigg]
\end{align}
We can add these two contributions and then separate the trace parts conveniently as
\begin{align}
\Xi^{\mu\nu\rho}_{(a+b)}&= -e^{3}\int\frac{d^{\omega}q}{(2\pi)^{\omega}}\frac{1}{(k^{2}-\mu^{2})(q^{2}-\mu^{2})(s^{2}-\mu^{2})}
\left(\delta^{\nu}_{\alpha}+\frac{m^{2}}{2}\frac{n_{\alpha} n^{\nu}}{(n.k)(n.q)}\right) \cr
&\times  \left(\delta^{\rho}_{\beta}+\frac{m^{2}}{2}\frac{n_{\beta} n^{\rho}}{(n.q)(n.s)}\right)\left(\delta^{\mu}_{\lambda}+\frac{m^{2}}{2}\frac{n_{\lambda} n^{\mu}}{(n.s)(n.k)}\right)\Big({\cal{A}}^{\alpha\beta\lambda}+{\cal{B}}^{\beta\alpha\lambda}\Big) \label{eq:20}
\end{align}
where we have defined
\begin{align}
{\cal{A}}^{\alpha\beta\lambda}&=\textrm{Tr}\Big[(\displaystyle{\not}\tilde{k}+m_{e})
\gamma^{\alpha} \left(\displaystyle{\not}\tilde{q}+m_{e}\right) \gamma^{\beta}
\left(\displaystyle{\not}\tilde{s}+m_{e}\right)
\gamma^{\lambda}
\Big], \cr
{\cal{B}}^{\beta\alpha\lambda}&=-\textrm{Tr}\Big[\left(\displaystyle{\not}\tilde{s}-m_{e}\right) \gamma^{\beta} \left(\displaystyle{\not}\tilde{q}-m_{e}\right) \gamma^{\alpha}
(\displaystyle{\not}\tilde{k}-m_{e}) \gamma^{\lambda} \Big].
\end{align}
We can show that every term, arising from these contributions, cancel mutually by straightforward manipulations using
\begin{equation}
\textrm{Tr}\Big(\gamma^{\mu_{1}}\gamma^{\mu_{2}}\ldots\gamma^{\mu_{n-1}}\gamma^{\mu_{n}}\Big)=
(-1)^{n}\textrm{Tr}\Big(\gamma^{\mu_{n}}\gamma^{\mu_{n-1}}\ldots\gamma^{\mu_{2}}\gamma^{\mu_{1}}\Big)
\end{equation}
that follows from the charge conjugation invariance, $C^{-1}\gamma^\mu C = - (\gamma^\mu)^T$ , and it is valid for any number of gamma matrices.
Hence, after some manipulations, we can show that
\begin{equation}
{\cal{B}}^{\beta\alpha\lambda} = - {\cal{A}}^{\alpha\beta\lambda}
\end{equation}
which implies that the triangle graphs (a) and (b) \eqref{eq:20} sum to zero
\begin{equation}
\Xi^{\mu\nu\rho}_{(a+b)} = 0 \label{eq:21}
\end{equation}
This result for VSR electrodynamics is the same as in the ordinary QED, and the nonlocal VSR contributions, in the fermionic propagator and couplings, do not change the outcome for the triangle graphs.
However, we have two additional graphs due to the VSR couplings for the one-loop vev $\langle AAA \rangle$, which must be explicitly evaluated to verify the validity of Furry's theorem at this order. Next, we shall show that each one of these two contributions vanishes independently.

We start from the graph (c), consisting of a diagram with the cubic and quartic vertex, which can be written as
\begin{align}
\Xi^{\mu\nu\rho}_{(c)}&=(-1)\int \frac{d^{\omega}q}{(2\pi)^{\omega}}\textrm{Tr} \Bigg[i\frac{\left(\displaystyle{\not}\tilde{q}+m_{e}\right)}{q^{2}-\mu^{2}}\Gamma^{\mu\nu}\left(q,u,p_{1},p_{2}\right)
i\frac{\left(\displaystyle{\not}\tilde{u}+m_{e}\right)}{u^{2}-\mu^{2}}\Lambda^{\rho}\left(q,u\right)\Bigg] \label{eq:22}
\end{align}
where, $u=q-p_{3}$. This expression can be simplified by computing the trace over the gamma matrices, by making use of the identity \eqref{eq:11} and Feynman parametrization.
Thus, after performing these manipulations, we are left with an expression that can be cast in a general form
as below
\begin{align}
\Xi_{(c)}^{\mu\nu\rho}(p_{1},p_{2},p_{3})	 &=a_{1}^{\mu\nu\rho}\left(m^{2},p_{i}\right)\left(\mathcal{J}_{1}^{\left(+\right)}+\mathcal{J}_{1}^{\left(-\right)}\right)+a_{2}^{\mu\nu\rho}\left(m^{2},p_{i}\right)\left(\mathcal{J}_{2}^{\left(+\right)}+\mathcal{J}_{2}^{\left(-\right)}\right) \cr
&	 +\left(a_{3}\right)_{\lambda}^{\mu\nu\rho}\left(m^{2},p_{i}\right)\left(\mathcal{J}_{3}^{\lambda\left(+\right)}-\mathcal{J}_{3}^{\lambda\left(-\right)}\right)+\left(a_{4}\right)_{\lambda}^{\mu\nu\rho}\left(m^{2},p_{i}\right)\left(\mathcal{J}_{4}^{\lambda\left(+\right)}-\mathcal{J}_{4}^{\lambda\left(-\right)}\right)\cr
&	 -a_{5}^{\mu\nu\rho}\left(m^{2},p_{i}\right)\left(\mathcal{J}_{5}^{\left(+\right)}+\mathcal{J}_{5}^{\left(-\right)}\right)+a_{5}^{\mu\nu\rho}\left(m^{2},p_{i}\right)\left(\mathcal{J}_{6}^{\left(+\right)}+\mathcal{J}_{6}^{\left(-\right)}\right) \label{eq:23}
\end{align}
where we have defined the quantities $a_{i}^{\mu\nu\rho}$ in the Appendix \ref{apB}, and also introduced the integrals $\mathcal{J}_{i}$ conveniently as
\begin{align}
\left(\mathcal{J}_{1}^{\left(\pm\right)},\mathcal{J}_{3}^{\lambda\left(\pm\right)},\mathcal{J}_{5}^{\left(\pm\right)}\right)	 &=\int dx\int\frac{d^{\omega}q}{(2\pi)^{\omega}}\frac{\left(1,q^{\lambda},q^{2}\right)}{\left(q^{2}\pm2xq. p_{1}-\Phi^{2}\right)^{2}}\frac{1}{n.q} \\
\left(\mathcal{J}_{2}^{\left(\pm\right)},\mathcal{J}_{4}^{\lambda\left(\pm\right)},\mathcal{J}_{6}^{\left(\pm\right)}\right)	 &=\int dx\int\frac{d^{\omega}q}{(2\pi)^{\omega}}\frac{\left(1,q^{\lambda},q^{2}\right)}{\left(q^{2}\pm2q\cdot(xp_{1}+p_{2})-\Omega^{2}\right)^{2}}\frac{1}{n. q}
\end{align}
in terms of
\begin{align}
\Phi^{2}	&=\mu^{2}-xp_{1}^{2}\\
\Omega^{2}	&=\mu^{2}-xp_{1}^{2}-2xp_{1}.p_{2}-p_{2}^{2}
\end{align}

Making use of the Mandelstam-Leibbrandt prescription, Eqs.~\eqref{eq:8}, \eqref{eq:9} and \eqref{eq:10}, one can show that
\begin{align}
\left(\mathcal{J}_{1}^{(+ )},\mathcal{J}_{3}^{\mu(+)},\mathcal{J}_{5}^{(+ )}\right) = \left( - \mathcal{J}_{1}^{(- )},\mathcal{J}_{3}^{\mu(-)}, -\mathcal{J}_{5}^{(- )}\right) \\
\left(\mathcal{J}_{2}^{(+ )},\mathcal{J}_{4}^{\mu(+)},\mathcal{J}_{6}^{(+ )}\right) = \left( - \mathcal{J}_{2}^{(- )},\mathcal{J}_{4}^{\mu(-)}, -\mathcal{J}_{6}^{(- )}\right)
\end{align}
Hence, this development shows that the whole contribution (c) vanishes, since every pair of terms $\mathcal{J}_{i}^{\left(\pm\right)}$ in the expression \eqref{eq:23} cancels itself exactly, implying that
\begin{equation}
\Xi^{\mu\nu\rho}_{(c)} = 0 \label{eq:24}
\end{equation}

The last piece that we shall discuss is the contribution arising from the graph (d), which is written as
\begin{equation}
\Xi_{(d)}^{\mu\nu\rho}=-\int  \frac{d^{\omega}q}{(2\pi)^{\omega}}\textrm{Tr}\Bigg[i\frac{\left(\displaystyle{\not}\tilde{q}+m_{e}\right)}
{q^{2}-\mu^{2}}\Gamma^{\mu\nu\rho}\left(q,q,p_{1},p_{2},p_{3}\right)\Bigg].
 \label{eq:25}
\end{equation}
After computing the trace of gamma matrices, we are left with
\begin{align}
\Xi_{(d)}^{\mu\nu\rho}&=e^{3}m^{2}n^{\mu}n^{\nu}n^{\rho}\int\frac{d^{\omega}q}{(2\pi)^{\omega}}\frac{1}{(q^{2}-\mu^{2})(n.q)}\cr
&\times\Bigg(\frac{1}{n.\left(q+p_{1}\right)n.\left(q-p_{3}\right)}+\frac{1}{n.\left(q+p_{1}\right)n.\left(q-p_{2}\right)}\cr
&+\frac{1}{n.\left(q+p_{2}\right)n.\left(q-p_{3}\right)}+\frac{1}{n.\left(q+p_{2}\right)n.\left(q-p_{1}\right)}\cr
&+\frac{1}{n.\left(q+p_{3}\right)n.\left(q-p_{2}\right)}+\frac{1}{n.\left(q+p_{3}\right)n.\left(q-p_{1}\right)}\Bigg).
\end{align}
In order to solve the momentum integrals, it is convenient to use again the identity \eqref{eq:11} to simplify the expression.
In this process, we make use of the result
\begin{equation}
\int \frac{d^{\omega}q}{(2\pi)^{\omega}}\frac{1}{(q^{2}-\mu^{2})(n.q)}=0,
\end{equation}
that follows from \eqref{eq:8}.
Thus, after some manipulations with \eqref{eq:11} and further simplifications, we finally arrive at the expression
\begin{align} \label{eq66}
\Xi_{(d)}^{\mu\nu\rho}&=\frac{e^{3}m^{2}n^{\mu}n^{\nu}n^{\rho}}{n.\left(p_{3}+p_{1}\right)n.\left(p_{3}+p_{2}\right)}\Big(\mathcal{K}_{3}^{(-)}+\mathcal{K}_{3}^{(+)}\Big)\cr
&+\frac{e^{3}m^{2}n^{\mu}n^{\nu}n^{\rho}}{n.\left(p_{2}+p_{1}\right)n.\left(p_{2}+p_{3}\right)}\Big(\mathcal{K}_{2}^{(-)}+\mathcal{K}_{2}^{(+)}\Big)\cr
&+\frac{e^{3}m^{2}n^{\mu}n^{\nu}n^{\rho}}{n.\left(p_{1}+p_{2}\right)n.\left(p_{1}+p_{3}\right)}
\Big(\mathcal{K}_{1}^{(-)}+\mathcal{K}_{1}^{(+)}\Big),
\end{align}
where we have introduced by simplicity the notation
\begin{equation}
\mathcal{K}_i^{(\pm)} = \int\frac{d^{\omega}q}{(2\pi)^{\omega}}\frac{1}{\left((q \pm p_{i})^{2}-\mu^{2}\right)\left(n.q\right)}, ~~i=1,2,3
\end{equation}
However, making use of the identity \eqref{eq:8}, we can show that
\begin{equation}
\mathcal{K}_{i}^{(-)} = - \mathcal{K}_{i}^{(+)}.
\end{equation}
This result, ultimately, implies that the contribution of the graph (d) \eqref{eq66} vanishes
\begin{equation}
\Xi_{(d)}^{\mu\nu\rho}=0.
\label{eq:26}
\end{equation}

In summary, from the results Eqs.~\eqref{eq:21}, \eqref{eq:24} and \eqref{eq:26}, we can conclude that the whole one-loop amplitude $\langle AAA \rangle$ is equal to zero
\begin{equation}
\Xi^{\mu\nu\rho} = \Xi_{(a+b)}^{\mu\nu\rho} + \Xi_{(c)}^{\mu\nu\rho}+\Xi_{(d)}^{\mu\nu\rho}= 0.
 \label{eq:27}
\end{equation}

This final result shows that, similarly to the ordinary QED, the VSR electrodynamics also satisfies Furry's theorem (at least in the one-loop order) and that no Chern-Simons-like self-coupling term is dynamically generated.
Although VSR changed the photon's dynamics in the free part of the Maxwell-Chern-Simons action, its Abelian structure and additional couplings are not sufficient to engender new self-couplings.
Of course, if we increase the number of external photon legs to four, i.e. the vev $\langle AAA A\rangle$, we could check the VSR contribution to the Euler-Heisenberg effective action in $(2+1)$-dim., similarly to the SIM(2) invariant analysis for a $(3+1)$ spacetime \cite{Alfaro:2020njh}.

\section{Final remarks}
\label{conc}

In this paper, we have discussed the photon's effective action in the context of VSR electrodynamics in the $(2+1)$ spacetime, with a special interest in analyzing the validity of Furry's theorem in the context of VSR.
Initially, we revised the main aspects regarding the VSR gauge symmetry, and how this invariance introduced a new covariant derivative, which implies an infinite series of nonlocal couplings among the fermionic and gauge fields.
After deriving the respective Feynman rules for the new VSR couplings, we proceeded to the evaluation of the respective graphs contributing to the one-loop amplitudes $\langle AA\rangle$ and $\langle AAA\rangle$, corresponding to the free and self-coupling parts of the photon's effective action.

In the analysis of the two-point function $\langle AA\rangle$, in addition to the usual polarization graph, there is a second diagram coming solely from the new VSR quartic coupling.
In order to solve the momentum integration, we used the Mandelstam-Leibbrant prescription extended to the VSR invariant case, where a new vector $\bar{n}$ is introduced \cite{Alfaro:2017umk}.
Additionally, to complete the analysis, it is necessary to determine an explicit form for this vector, where some properties are considered: reality, symmetry, etc, which ultimately implied in a form preserving all of these features.

Finally, we considered the low-momentum limit in order to determine the dynamical part of the effective action.
There, we have obtained the usual Maxwell-Chern-Simons terms, added by VSR contributions both to the parity even and parity odd sectors.
 However, the VSR effects are different in the classical and quantum realm, since the nonlocal and higher-derivative terms are intermingled in the quantum effective action.
This entanglement of nonlocal effects and higher-derivative terms, in the gauge sector, signals that we have the presence of UV/IR mixing in the VSR quantum effective action.
Furthermore, we have discussed possible applications of the VSR-Chern-Simons effective action in topological invariant systems, such as quantum Hall systems and other topological insulators.
In particular, we have analyzed the VSR contribution to Hall's conductivity, and showed that it vanishes.
However, Lorentz violating effects can be well motivated in the context of topological invariant systems in order to describe anomalous behavior, such as non-Ohmic materials.

The last piece of the effective action that we have analyzed was the three-point function $\langle AAA\rangle$. In this case, we had the same triangle graphs from ordinary QED, but due to the VSR couplings, two new graphs contributed to the complete one-loop amplitude.
On one side, the two triangle graphs canceled mutually by using simply properties of the gamma matrices trace, which are based in the charge conjugation symmetry.
Actually, this result for the triangle graphs is independent of the VSR coupling, i.e. the parameter $m^2$, since $\Xi_{(a+b)}^{\mu\nu\rho}$ has the same matrix structure of the ordinary QED.
In regard to the two additional graphs, due to the presence of the VSR couplings, we explicitly showed that these contributions vanish individually.
Hence, based on the fact that the amplitude $\langle AAA\rangle$ vanished, we verified that Furry's theorem is satisfied (at least in the one-loop order), and that no Chern-Simons-like self-coupling term is dynamically generated.
Although VSR changed the photon's dynamics in the Maxwell-Chern-Simons action, its Abelian structure and additional couplings are not sufficient to engender new self-couplings.

Since the induced one-loop effective action for the non-Abelian gauge fields interacting with Dirac fermions in $d=3$ leads to the known Yang-Mills (even-parity) and non-Abelian Chern-Simons (odd-parity) terms, it would be interesting to generalize our analysis to the case of a non-Abelian VSR gauge theory \cite{Alfaro:2013uva,Alfaro:2015fha}.
This analysis would allow us to observe, in particular, how the odd-parity terms change under VSR effects, possibly resulting in a different type of self-coupling terms \cite{prep1}.
Another point of interest is the study of the general tensorial structure of the photon self-energy in different spacetime dimensions ($d=3,4$) at any order of perturbation, leading to the full photon propagator.
This study permits us to investigate the physical pole structure of the photon propagator and also to analyze  possible VSR corrections to the topological mass in the Chern-Simons theory \cite{prep2}.

 \subsection*{Acknowledgements}

The authors would like to thank the anonymous referee for his/her comments and suggestions to improve this paper. We are also grateful to M.M. Sheikh-Jabbari for his comments on the manuscript.  R.B. acknowledges partial support from Conselho
Nacional de Desenvolvimento Cient\'ifico e Tecnol\'ogico (CNPq Projects No. 305427/2019-9 and No. 421886/2018-8) and Funda\c{c}\~ao de
Amparo \`a Pesquisa do Estado de Minas Gerais (FAPEMIG Project No. APQ-01142-17).

\appendix

\section{Useful integrals}
\label{apA}

In order to cope with the momentum integral in VSR, we use the Mandelstam-Leibbrant prescription, and further useful results can be obtained from \eqref{eq:8} by taking a derivative in relation to $p_\mu$
\begin{align}
\int d^{\omega}q & \frac{q_{\mu}}{(q^2 + 2 q.p - m^2)^{a + 1}} \frac{1}{(n
\cdot q)^b} \cr
& = (- 1)^{a + b} i \pi^{\frac{\omega}{2}} (- 2)^{b - 1} \frac{\Gamma (a + b -
\frac{\omega}{2})}{\Gamma (a + 1) \Gamma (b)} (\bar{n}.p)^{b - 1} b
\bar{n}_{\mu} \int^1_0 d t t^{b - 1} \frac{1}{\Delta^{a + b - \frac{\omega}{2}}}  \cr
&+(- 1)^{a + b} i \pi^{\frac{\omega}{2}} (- 2)^b \frac{\Gamma (a + b + 1 -
\frac{\omega}{2})}{\Gamma (a + 1) \Gamma (b)} (\bar{n}.p)^b  \int^1_0 d t t^{b -
1} \frac{p_{\mu} - t (n.p \bar{n}_{\mu} + \bar{n}.p
n_{\mu})}{\Delta^{a + b + 1 - \frac{\omega}{2}}} \label{eq:9}
\end{align}
where  $\Delta = m^2 + p^2 - 2 (n \cdot p) (\bar{n} \cdot p) t$, another useful result
is found by taking a derivative of \eqref{eq:9} in relation to $p_\nu$ and contracting with the metric, implying
\begin{align}
\int d^{\omega}q & \frac{q^2}{(q^2 + 2 q \cdot p - m^2)^{a + 2}} \frac{1}{(n
\cdot q)^b} = \cr
=& (- 1)^{a + b} i \pi^{\frac{\omega}{2}} (- 2)^{b - 2} \bigg\{ -4 \frac{\Gamma (a + b +
1 - \frac{\omega}{2})}{\Gamma (a + 2) \Gamma (b)} (\bar{n}. p)^b b \int^1_0 d t
t^{b - 1} \frac{(1 - t)}{\Delta^{a + b + 1 - \frac{\omega}{2}}} \cr
& + 4
\frac{\Gamma (a + b + 2 - \frac{\omega}{2})}{\Gamma (a + 2) \Gamma (b)} (\bar{n}.
p)^b  \int^1_0 d t t^{b - 1} \frac{[p^2 - 4 t (n \cdot p \bar{n}.p) +
2 t^2 (n.p \bar{n}.p)]}{\Delta^{a + b + 2 - \frac{\omega}{2}}} \cr
&- 2 \frac{\Gamma (a + b + 1 -
\frac{\omega}{2})}{\Gamma (a + 2) \Gamma (b)} (\bar{n}.p)^b \int^1_0 d t t^{b -
1} \frac{\frac{\omega}{2} - 2 t}{\Delta^{a + b +
1 - \frac{\omega}{2}}} \bigg\} \label{eq:10}
\end{align}

Furthermore, in the evaluation of vev $\langle AA \rangle$ in section \ref{sec3} we have defined some integrals by simplicity of notation
\begin{align}
\mathcal{I}_1=&\int_{0}^{1} dx~\frac{x(1-x)}{\sqrt{\mu^{2}-x(1-x)p^{2}} }=-
\frac{\sqrt{\mu^2}}{2 p^2} + \frac{(4 \mu^2 + p^2)}{(p^2)^{3 / 2}} \ln
\left(\frac{ 2 \sqrt{\mu^2 p^2} + p^2}{ 2 \sqrt{\mu^2 p^2} - p^2} \right) \label{int_1} \\
\mathcal{I}_2=&\int_{0}^{1}dx~\frac{1}{\sqrt{\mu^{2}-x(1-x)p^{2}}} = \frac{1}{\sqrt{p^2}}\ln
\left(\frac{ 2 \sqrt{\mu^2 p^2} + p^2}{ 2 \sqrt{\mu^2 p^2} - p^2} \right) \label{int_2} \\
\mathcal{I}_3=&\int_{0}^{1} dx~\frac{1}{x}\Bigg(\frac{1}{\sqrt{\mu^{2}-xp^{2}}}-\frac{1}{\sqrt{\mu^{2}-x(1-x)p^{2}}}\Bigg) \cr
 =&\frac{1}{\sqrt{\mu^{2}}}\left[\ln\left(\frac{4\mu^{2}}{p^{2}}-1\right)-\ln\left(\frac{1+\sqrt{1-\frac{p^{2}}{\mu^{2}}}}{1-\sqrt{1-\frac{p^{2}}{\mu^{2}}}}\right)\right] \label{int_3}
\end{align}
Afterwards, we have considered the low-momentum limit of these expressions in  \eqref{eq:16} to determine the dynamical part of the photon's effective action.

\section{Tensor quantities}
\label{apB}

We present here some tensor quantities we have introduced in the evaluation of one-loop contribution for the vev $\langle AAA \rangle$ in section \ref{sec4}.
In particular, we have the tensor quantities present in the graph (c) expression
\begin{align}
a_{1}^{\mu\nu\rho}\left(m^{2},p_{i}\right)	&=\frac{m^{2}e^{3}\mu^{2}n^{\mu}n^{\nu}n^{\rho}}{\left(n. p_{2}\right)n.\left(p_{1}+p_{2}\right)}+\frac{im^{2}m_{e}e^{3}\varepsilon^{\mu\alpha\beta}p_{1\alpha}n_{\beta}n^{\nu}n^{\rho}}{\left(n. p_{2}\right)n.\left(p_{1}+p_{2}\right)} \\
a_{2}^{\mu\nu\rho}\left(m^{2},p_{i}\right)	&=-\frac{m^{2}e^{3}\mu^{2}n^{\mu}n^{\nu}n^{\rho}}{\left(n. p_{2}\right)n.\left(p_{1}+p_{2}\right)}+\frac{im^{2}m_{e}e^{3}\varepsilon^{\mu\alpha\beta}p_{1\alpha}n_{\beta}n^{\nu}n^{\rho}}{\left(n. p_{2}\right)n.\left(p_{1}+p_{2}\right)}\cr
&	-\frac{m^{2}e^{3}n^{\nu}n^{\rho}}{\left(n. p_{2}\right)n.\left(p_{1}+p_{2}\right)}\Big(n.\left(2p_{2}+p_{1}\right)p_{2}^{\mu}-p_{2}.\left(p_{1}+p_{2}\right)n^{\mu}+\left(n. p_{2}\right)p_{1}^{\mu}\Big)\\
\left(a_{3}\right)_{\lambda}^{\mu\nu\rho}\left(m^{2},p_{i}\right)&	=\frac{m^{2}e^{3}n^{\nu}n^{\rho}\left(\left(n. p_{1}\right)\delta_{\lambda}^{\mu}-\left(p_{1}\right)_{\lambda}n^{\mu}\right)}{\left(n. p_{2}\right)n.\left(p_{1}+p_{2}\right)} \\
\left(a_{4}\right)_{\lambda}^{\mu\nu\rho}\left(m^{2},p_{i}\right)	&=\frac{m^{2}e^{3}n^{\nu}n^{\rho}}{\left(n. p_{2}\right)n.\left(p_{1}+p_{2}\right)}\Big(\left(2p_{2}+p_{1}\right)_{\lambda}n^{\mu}-n.\left(2p_{2}+p_{1}\right)\delta_{\lambda}^{\mu}\Big)\\
a_{5}^{\mu\nu\rho}\left(m^{2},p_{i}\right)	&=\frac{m^{2}e^{3}n^{\mu}n^{\nu}n^{\rho}}{\left(n. p_{2}\right)n.\left(p_{1}+p_{2}\right)}
\end{align}


\global\long\def\link#1#2{\href{http://eudml.org/#1}{#2}}
 \global\long\def\doi#1#2{\href{http://dx.doi.org/#1}{#2}}
 \global\long\def\arXiv#1#2{\href{http://arxiv.org/abs/#1}{arXiv:#1 [#2]}}
 \global\long\def\arXivOld#1{\href{http://arxiv.org/abs/#1}{arXiv:#1}}


\end{document}